\documentclass[a4paper]{jpconf}
\usepackage{graphicx}
\begin{document}
\title{Spontaneous order in the highly frustrated spin-1/2 Ising-Heisenberg model 
on the triangulated Kagom\'e lattice due to the Dzyaloshinskii-Moriya anisotropy}

\author{J Stre\v{c}ka and L \v{C}anov\'a}

\address{Department of Theoretical Physics and Astrophysics, Faculty of Science, \\ 
P. J. \v{S}af\'{a}rik University, Park Angelinum 9, 040 01 Ko\v{s}ice, Slovak Republic}
\ead{jozef.strecka@upjs.sk}

\begin{abstract}
The spin-1/2 Ising-Heisenberg model on the triangulated Kagom\'e (triangles-in-triangles) 
lattice is exactly solved by establishing a precise mapping correspondence to the simple 
spin-1/2 Ising model on Kagom\'e lattice. It is shown that the disordered spin liquid state, 
which otherwise occurs in the ground state of this frustrated spin system on assumption that 
there is a sufficiently strong antiferromagnetic intra-trimer interaction, is eliminated 
from the ground state by arbitrary but non-zero Dzyaloshinskii-Moriya anisotropy.
\end{abstract}

\section{Introduction}

\begin{figure}[h]
\begin{center}
\includegraphics[width=12cm]{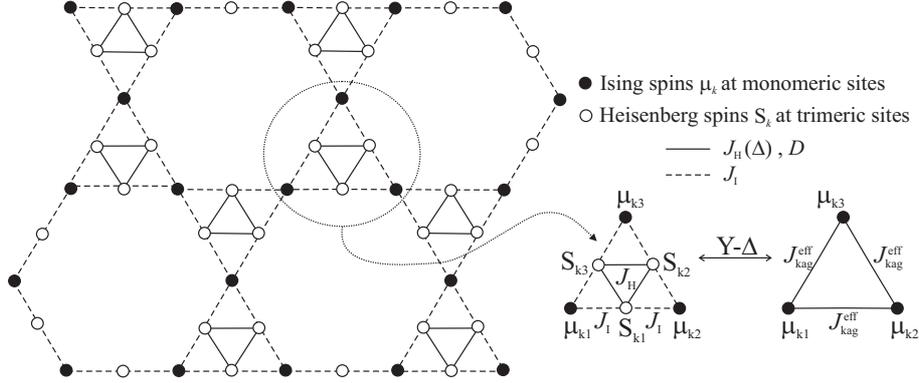}
\end{center}
\vspace{-0.5cm}
\caption{The cross-section of the triangulated Kagom\'e lattice. Open and full circles denote 
lattice positions of the Heisenberg and Ising spins. Solid lines denote the intra-trimer Heisenberg interaction $J_{\rm H} (\Delta)$ and Dzyaloshinskii-Moriya interaction $D$, while broken lines label the monomer-trimer Ising interaction $J_{\rm I}$. The ellipse marks $k$th Heisenberg trimer and its three enclosing Ising spins described by the Hamiltonian (\ref{eq1}), which can be mapped via the star-triangle transformation to a simple triangle of the Ising spins coupled by the effective interaction $J_{\rm kag}^{\rm eff}$.}
\label{fig1}
\end{figure}

The antiferromagnetic quantum Heisenberg model (QHM) on \textit{geometrically frustrated planar lattices} is currently at a forefront of theoretical research interest as it exhibits a variety 
of unusual ground states owing to a mutual interplay between quantum fluctuations and geometric frustration \cite{Lhu02,Mis04,Ric04}. The intensive efforts aimed at better understanding of 
this peculiar interplay are closely related to several transition-metal magnetic materials, which are topologically prone to the geometric frustration due to a connectivity of their underlying magnetic lattice \cite{Gre01,Har04}. One of the most interesting geometrically frustrated magnetic structures 
is the triangulated Kagom\'e (triangles-in-triangles) lattice (Fig.~\ref{fig1}), which resembles the magnetic structure of a series of polymeric coordination compounds Cu$_9$X$_2$(cpa)$_6$.nH$_2$O 
(X=F, Cl, Br) \cite{Nor87,Nor90,Gon93}. 

It is worthy to notice, however, that the spin-1/2 QHM on the triangulated Kagom\'e lattice is 
not amenable to exact analytical treatment on account of mathematical complexities closely connected
to a non-commutability of spin operators. Contrary to this, the spin-1/2 Ising model on the triangulated Kagom\'e lattice can be rather straightforwardly solved by establishing a precise 
mapping relationship to the spin-1/2 Ising model on Kagom\'e \cite{Zhe05} or honeycomb \cite{Loh08} lattice exactly solved many years ago. The Ising model might unfortunately fail in describing many important features of the copper-based coordination compounds as it entirely neglects quantum fluctuations firmly associated with a quantum nature of the paramagnetic Cu$^{2+}$ ions having 
the lowest possible quantum spin number $1/2$. Nevertheless, it has been recently shown that the approach based on exact mapping relations can also be applied to the spin-1/2 Ising-Heisenberg model 
on triangulated Kagom\'e lattice, which correctly takes into account the intra-trimer Heisenberg interaction and approximates merely the monomer-trimer interaction by the Ising-type coupling \cite{Str08,Yao08}. In the present work, we shall provide a further extension of this model by accounting for the antisymmetric Dzyaloshinskii-Moriya interaction as well. 

The rest of this paper is so organized. In Section 2, we briefly describe the investigated model 
system and we also recall the most important steps of exact mapping method. Section 3 deals with 
a discussion of the most interesting results and it also summarizes main conclusions.   

\section{Ising-Heisenberg model and its exact solution}

Consider the spin-1/2 Ising-Heisenberg model on the triangulated Kagom\'e lattice, 
which consists of two different lattice sites diagrammatically depicted in Fig.~\ref{fig1} 
as open and full circles. Assign the \textit{Heisenberg spin} $S=1/2$ to all lattice sites schematically shown as open circles and the \textit{Ising spin} $\mu=1/2$ to all lattice 
sites shown as full circles. The spin-1/2 Ising-Heisenberg model on the triangulated Kagom\'e 
lattice can be then viewed as the Kagom\'e lattice of the Ising spins (monomers), which contains 
inside of each triangle unit a smaller triangle of the Heisenberg spins (trimer). The total 
Hamiltonian of the model under investigation can be for further convenience written as a sum over 
all Heisenberg trimers $\hat{{\mathcal H}} = \sum_{k} \hat{{\mathcal H}}_k$, where each Hamiltonian $\hat{{\mathcal H}}_k$ involves all the interaction terms associated with three Heisenberg spins 
from $k$th trimer
\begin{eqnarray}
\hat{{\mathcal H}}_k = \sum_{i=1}^3  \biggl \{
\! \! \! \! &-& \! \! \! \! J_{\rm H} 
  \left[\Delta \left(\hat{S}_{k,i}^x \hat{S}_{k,i+1}^x + \hat{S}_{k,i}^y \hat{S}_{k,i+1}^y \right) 
           + \hat{S}_{k,i}^z \hat{S}_{k,i+1}^z \right] \nonumber \\
        \! \! \! \! &-& \! \! \! \! D         
        \left(\hat{S}_{k,i}^x \hat{S}_{k,i+1}^y - \hat{S}_{k,i}^y \hat{S}_{k,i+1}^x \right) 
        - J_{\rm I} \hat{S}_{k,i}^z \left(\hat{\mu}_{k,i}^z + \hat{\mu}_{k,i+1}^z \right) \biggr \}.
\label{eq1}	                        
\end{eqnarray}
The parameter $J_{\rm H}$ denotes the intra-trimer Heisenberg interaction, $\Delta$ is the spatial anisotropy in this interaction, $D$ labels the intra-trimer Dzyaloshinskii-Moriya interaction, 
$J_{\rm I}$ stands for the monomer-trimer Ising interaction and $\hat{S}_{k,4} \equiv \hat{S}_{k,1}$, $\hat{\mu}_{k,4} \equiv \hat{\mu}_{k,1}$. 

Exact solution for the aforedescribed model system can be achieved by following the same procedure 
as developed in our earlier work \cite{Str08}. In fact, the exact mapping correspondence 
to the simple spin-1/2 Ising model on Kagom\'e lattice can be established by adopting 
the star-triangle transformation given by Eq.~(4) of Ref.~\cite{Str08} with the modified 
parameters $P$ and $Q_{\pm}$
\begin{eqnarray}
P \! \! \! &=& \! \! \! \left(\frac{J_{\rm I}}{3} \right)^2 \left[\frac{3}{4} - (\mu_{k1}^z \mu_{k2}^z 
             + \mu_{k2}^z \mu_{k3}^z  + \mu_{k3}^z \mu_{k1}^z) \right] 
             + \left(\frac{J_{\rm H} \Delta}{2} \right)^2 + \left(\frac{D}{2} \right)^2 \! \! \!,  
             \label{eq2a}	\\
Q_{\pm} \! \! \! &=& \! \! \! \pm \frac{1}{2} \left(\frac{J_{\rm I}}{3} \right)^3 \left[\mu_{k1}^z  
         + \mu_{k2}^z + \mu_{k3}^z - 12 \mu_{k1}^z \mu_{k2}^z \mu_{k3}^z \right] 
         - \left(\frac{J_{\rm H} \Delta}{2} \right) 
      \left[ \left(\frac{J_{\rm H} \Delta}{2} \right)^2 - 3 \left(\frac{D}{2} \right)^2 \right]\!. \label{eq2b}	
\end{eqnarray}
In the spirit of this transformation, the spin-1/2 Ising-Heisenberg model on the triangulated Kagom\'e lattice is mapped to the corresponding spin-1/2 Ising model on the simple Kagom\'e lattice with 
the effective nearest-neighbour interaction $\beta J_{\rm kag}^{\rm eff} = \ln(V_1/V_2)$ given by
\begin{eqnarray}
V_1 \! \! \! &=& \! \! \! 
              2 \exp \left(\beta J_{\rm H} \right) \cosh \left(\frac{3\beta J_{\rm I}}{2} \right) 
            + 2 \cosh \left(\frac{\beta J_{\rm I}}{2} \right)
                \sum_{n=0}^2 \exp \left[-2 \beta \mbox{sgn}(q_{1}) \sqrt{p_1} 
                \cos \left(\phi_{1} + \frac{2 \pi n}{3} \right) \right]\! \!, \label{eq3a} \\
V_2 \! \! \! &=& \! \! \! 
              2 \exp \left(\beta J_{\rm H} \right) \cosh \left(\frac{\beta J_{\rm I}}{2} \right) 
            + \exp \left(\frac{\beta J_{\rm I}}{6} \right) \sum_{n=0}^2 
              \exp \left[-2 \beta \mbox{sgn}(q_2^{+}) \sqrt{p_2} \cos \left(\phi_2^{+} 
                    + \frac{2 \pi n}{3} \right) \right]  \nonumber \\
\! \! \! &+& \! \! \! \exp \left(- \frac{\beta J_{\rm I}}{6} \right) \sum_{n=0}^2 
    \exp \left[-2 \beta \mbox{sgn}(q_2^{-}) \sqrt{p_2} \cos \left(\phi_2^{-} 
                   + \frac{2 \pi n}{3} \right) \right]\! \!, 
\label{eq3b}	
\end{eqnarray}
where $\beta = 1/(k_{\rm B} T)$, $k_{\rm B}$ is Boltzmann's constant, $T$ absolute temperature 
and the parameters
\begin{eqnarray}
p_1 \! \! \! &=& \! \! \! 
  \left(\frac{J_{\rm H} \Delta}{2} \right)^2 \! \! + \left(\frac{D}{2} \right)^2\! \! \! \!, \, \, \,
q_{1} = \left(\frac{J_{\rm H} \Delta}{2} \right) \left[3 \left(\frac{D}{2}\right)^2 \! \!  
       - \left(\frac{J_{\rm H} \Delta}{2} \right)^2 \right]\! \!, \, \, \, 
\phi_{1} = \frac{1}{3} \arctan \left(\frac{\sqrt{p_1^3 - q_{1}^2}}{q_{1}} \right)\! \!; \nonumber \\
p_2 \! \! \! &=& \! \! \! p_1 + \left(\frac{J_{\rm I}}{3} \right)^2\! \! \! \!, \qquad \quad
q_2^{\pm} =  q_1 \pm \left(\frac{J_{\rm I}}{3} \right)^3\! \! \! \!, \qquad \quad
\phi_{2}^{\pm} = \frac{1}{3} \arctan \left(\frac{\sqrt{p_2^3 - (q_{2}^{\pm})^2}}{q_{2}^{\pm}} 
\right)\! \!. \label{eq4}	
\end{eqnarray}  
As a result of this mapping, the spin-1/2 Ising-Heisenberg model on the triangulated Kagom\'e lattice becomes critical if and only if the corresponding spin-1/2 Ising model on the Kagom\'e lattice becomes critical as well. Consequently, the critical temperature between ordered and disordered states of the Ising-Heisenberg model can be obtained by solving numerically the critical condition $\beta_{c} J_{\rm kag}^{\rm eff} = \ln (3 + 2\sqrt{3})$, where $\beta = 1/(k_{\rm B} T_{\rm c})$ and $T_{\rm c}$ is the critical temperature.

\section{Results and discussion}

Let us turn our attention to a discussion of the most interesting numerical results obtained 
for the finite-temperature phase diagrams. Before starting our discussion, however, it is worthwhile 
to remark that the most interesting results for the Ising-Heisenberg model without the Dzyaloshinskii-Moriya term has been detailed examined in our preceding paper \cite{Str08} 
to which the interested reader is referred to for more details. In the present work, 
we therefore aim to investigate mostly the effect of Dzyaloshinskii-Moriya interaction 
on the critical behaviour. 

\begin{figure}[h]
\begin{minipage}{7.5cm}
\includegraphics[width=7.5cm]{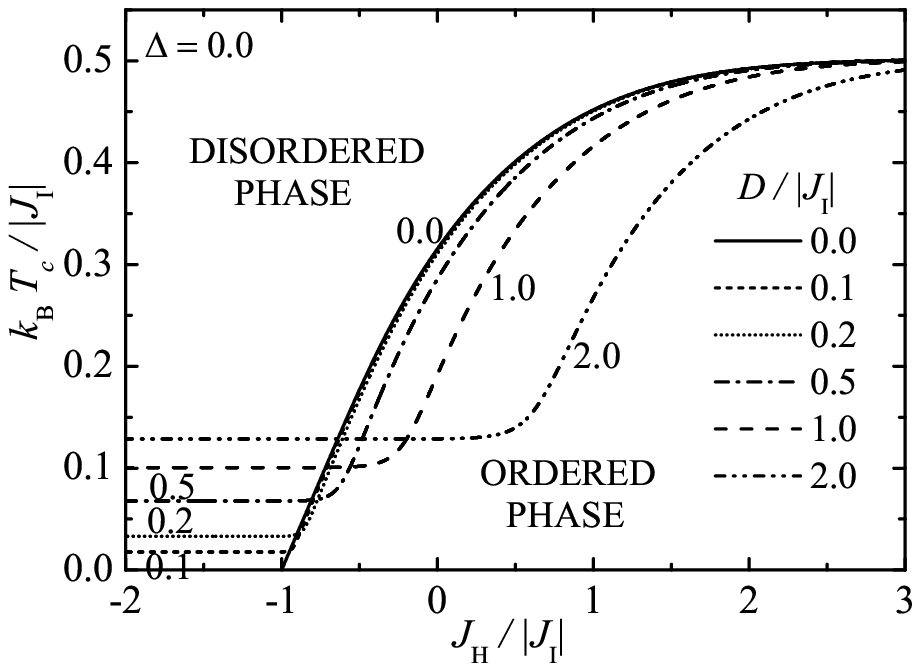}
\vspace{-1.3cm}
\caption{\label{label}The dimensionless critical temperature as a function of the ratio
$J_{\rm H}/|J_{\rm I}|$ for $\Delta = 0.0$ and several values of $D/|J_{\rm I}|$.}
\label{fig2}
\end{minipage}\hspace{2pc}%
\begin{minipage}{7.5cm}
\includegraphics[width=7.5cm]{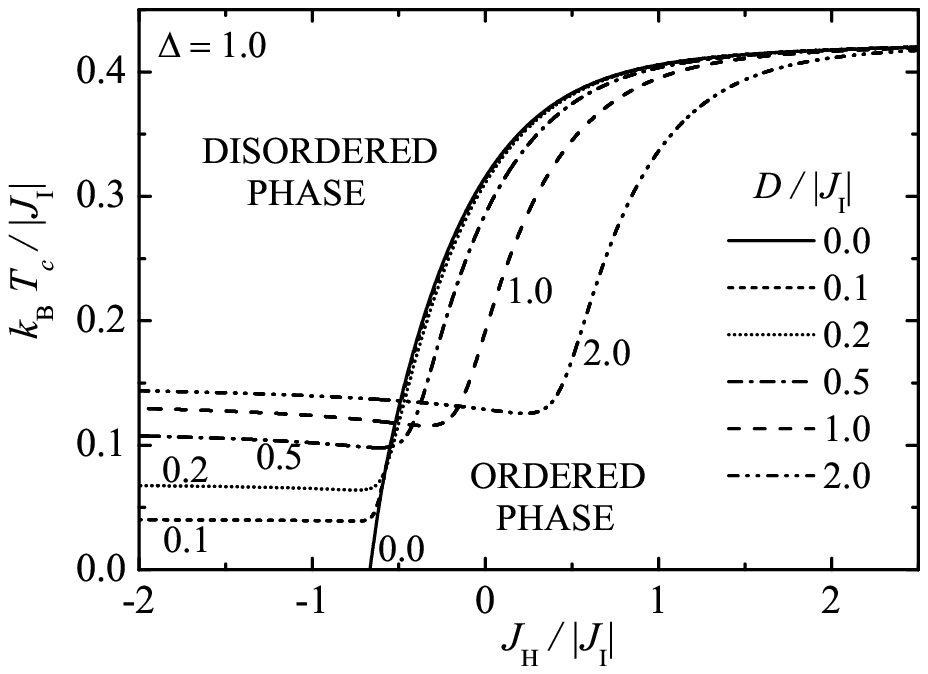}
\vspace{-1.3cm}
\caption{\label{label}The dimensionless critical temperature as a function of the ratio
$J_{\rm H}/|J_{\rm I}|$ for $\Delta = 1.0$ and several values of $D/|J_{\rm I}|$.}
\label{fig3}
\end{minipage} 
\end{figure}

In Figs.~\ref{fig2} and \ref{fig3}, the dimensionless critical temperature 
$k_{\rm B} T_{\rm c}/|J_{\rm I}|$ is plotted against the ratio $J_{\rm H}/|J_{\rm I}|$ between 
intra-trimer and monomer-trimer interactions for several values of the Dzyaloshinskii-Moriya 
term $D/|J_{\rm I}|$ and two different values of the exchange anisotropy $\Delta = 0.0$ 
and $1.0$, respectively. As it can be clearly seen, the displayed critical lines terminate 
at a certain value of the ratio $J_{\rm H}/|J_{\rm I}|$, below which the system becomes 
disordered at all temperatures, just when the Dzyaloshinskii-Moriya anisotropy completely vanishes 
(i.e. for $D/|J_{\rm I}| = 0.0$). 
This observation is fully consistent with previously published results \cite{Str08,Yao08}, 
which serve in evidence that the ground state of the spin-1/2 Ising-Heisenberg model 
without the Dzyaloshinskii-Moriya term is the spontaneously long-range ordered unless 
the geometric frustration is strong enough to raise the disordered spin liquid ground state for 
$J_{\rm H}/|J_{\rm I}| < - 2/(2 + \Delta)$. On the other hand, it surprisingly turns out 
that the ground state of the Ising-Heisenberg model with arbitrary but non-zero Dzyaloshinskii-Moriya 
anisotropy is spontaneously long-range ordered independently of $J_{\rm H}/|J_{\rm I}|$ and $\Delta$. 
As a matter of fact, the critical temperature tends asymptotically to some constant non-zero 
value for any $D/|J_{\rm I}| \neq 0.0$ even for highly frustrated case to be achieved in the 
limit $J_{\rm H}/|J_{\rm I}| \to -\infty$, whereas the asymptotical value of critical temperature 
is being the greater, the stronger is the Dzyaloshinskii-Moriya anisotropy $D/|J_{\rm I}|$. 
It should be also mentioned that the aforedescribed critical behaviour represents a rather 
general feature of the considered model that holds irrespective of the exchange anisotropy 
$\Delta$. Another striking observation for the highly frustrated region ($J_{\rm H}/|J_{\rm I}| \ll 0.0$) follows from a direct comparison of Figs.~\ref{fig2} and \ref{fig3}. It seems that 
quantum fluctuations support the spontaneous ordering, which is demonstrated by an increase 
of the critical temperature acquired through an increase of the parameter $\Delta$.

In conclusion, we have studied the spin-1/2 Ising-Heisenberg model on the triangulated Kagom\'e 
lattice within an exact method based on the generalized star-triangle mapping transformation. 
The most interesting result to emerge from our study is that the Dzyaloshinskii-Moriya anisotropy
prohibits an existence of the disordered spin liquid ground state in spite of the high geometric frustration caused by the antiferromagnetic intra-trimer interaction. The more comprehensive investigations of spontaneously ordered state(s) to appear in the highly frustrated region 
and of thermodynamics are left as challenging tasks for future work.

\ack{This work was supported by the Slovak Research and Development Agency under the contract 
LPP-0107-06 and by Ministry of Education of SR under the grant No.~VEGA~1/0128/08.}

\end{document}